\documentclass[12pt]{article}
\usepackage{graphics}
\def\ben{\begin{equation}}
\def\een{\end{equation}}
\def\bey{\begin{eqnarray}}
\def\eey{\end{eqnarray}}
\def\ba{\begin{array}}
\def\ea{\end{array}}

\def\lgl{\langle}
\def\rgl{\rangle}

\def\psla{p{\raise1pt\hbox{$\!\!/$}}}
\def\dsla{\partial{\raise1pt\hbox{$\!\!\!/$}}}
\def\Dsla{D{\raise1pt\hbox{$\!\!\!/$}}}
\def\xsla{x{\raise1pt\hbox{$\!\!\!/$}}}

\def\jmu5{j_{\mu 5}^{(i)}(0)}
\def\jnu5{j_{\nu 5}^{(i)}(0)}

\def\gA#1{g_{A}^{({#1})}(0)}
\def\ga#1{g_{A}^{({#1})}(q^2)}
\def\ha#1{h_{A}^{({#1})}(q^2)}

\def\qqv{\langle{\bar q}q\rangle_{0}}
\def\uuv{\langle{\bar u}u\rangle_{0}}

\def\ddv{\langle{\bar d}d\rangle_{0}}

\def\ssv{\langle{\bar s}s\rangle_{0}}

\def\qq0v{\langle0\!\mid\!{\bar q}q\!\mid\! 0\rangle}

\def\qc0f{\langle0\!\mid\!{\bar q}q\!\mid\!0\rangle_{F}}

\def\qsq0f{\langle0\!\mid\!{\bar q}\sigma_{\mu\nu}q\!\mid\!0\rangle_{F}}

\def\qgdq0f{\langle0\!\mid\!{\bar q}{\cal 
S}\gamma_{\mu}D_{\nu}q\!\mid\!0\rangle_{F}}

\def\qddq0f{\langle0\!\mid\!{\bar q}{\cal
S}D_{\mu}D_{\nu}q\!\mid\!0\rangle_{F}}
\def\m#1{m_{#1}}
\def\N#1{\langle{\bar {#1}}{#1}\rangle_{N}}
\def\ggN{\left\langle\frac{\alpha_s}{\pi}G^2\right\rangle_{N}}
\def\gdN#1{i\left\langle{\bar {#1}}{\cal 
S}\gamma_{\mu}D_{\nu}{#1}\right\rangle_{N}}
\def\gdddN#1{i\left\langle{\bar {#1}}{\cal 
S}\gamma_{\mu}D_{\nu}D_{\lambda}D_{\sigma}{#1}\right\rangle_{N}}
\def\sgmtrm{\Sigma_{\pi N}}
\def\mmnt#1#2{A_{#1}^{#2}(\mu^2)}
\def\mmntQSR#1#2{A_{#1}^{#2}(1\,{\rm GeV}^2)}
\def\3mmtm{|{\bf q}|^2}

\def\eq#1{Eq.(\ref{#1})}
\def\eqs#1#2{Eqs.(\ref{#1}) and (\ref{#2})}

\def\Ref#1{[\ref{#1}]}
\def\Refs#1#2{[\ref{#1},\ref{#2}]}

\def\nnbr{\nonumber}
\def\p0{p_0}

\def\gam3{\mbox{\boldmath{$\gamma$}}}
\def\e0{E_{0}(s_{0},s)}
\def\e1{E_{1}(s_{0},s)}
\def\e2{E_{2}(s_{0},s)}

\def\aplt{\kern0.3333em \raise 0.2ex \hbox{$<$}%
\kern-0.8em \lower0.8ex \hbox{$\sim$}%
\kern0.3333em}
\begin{document}


\baselineskip 30pt
\begin{center}
{\LARGE\bf A new approach to axial coupling constants in the QCD sum rule}
\end{center}

\vspace{.5cm}

\begin{center}
{\large Tetsuo NISHIKAWA
\footnote{E-mail address nishi@nuc-th.phys.nagoya-u.ac.jp}}
and
{\large Sakae SAITO
\footnote{E-mail address saito@nuc-th.phys.nagoya-u.ac.jp} 
}\\
\vspace{.2cm}
{\it Department of Physics, Nagoya University, Nagoya 464-8602, Japan}

\vspace{1.0cm}
{\large Yoshihiko KONDO\footnote{E-mail address kondo@kokugakuin.ac.jp}}\\
\vspace{0.2cm}
{\it Kokugakuin University, Tokyo 150-8440, Japan}
\end{center}


\baselineskip 18pt

\noindent{\bf Abstract}\par
{We derive new QCD sum rules for the axial coupling
constants by considering two-point correlation functions of the
axial-vector currents in a one-nucleon state. 
The QCD sum rules tell us that the axial coupling constants are expressed by 
nucleon matrix elements of quark and gluon operators 
which are related to the sigma-terms and the moments of parton distributions.
The results for the iso-vector axial coupling 
constants and the 8th component of the SU(3)
octet are in good agreement with experiment.}

\newpage
\setcounter{equation}{0}
The axial coupling constants are defined by the nucleon matrix 
elements of the axial-vector currents at zero momentum transfer.
The iso-vector axial coupling constant $\gA{3}$ and 
the 8th component of the SU(3) octet $\gA{8}$
are known to be expressed by the SU(3) parameters $F$ and $D$: $\gA{3}=F+D$, 
$\gA{8}=3F-D$. In the naive parton model
the singlet axial coupling constant $\gA{0}$ is expressed by the
fraction of the nucleon spin carried by the $q$ quark, $\Delta q$, as
$\gA{0}=\Delta u+\Delta d+\Delta s$, while $\gA{3}=\Delta u-\Delta d$, and
$\gA{8}=\Delta u+\Delta d-2\Delta s$. 
These coupling constants yield an important information on 
the spin structure 
of the valence and the sea quarks in the nucleon.
The parameters $F$ and $D$ are precisely known from the measurements 
in neutron and hyperon $\beta$-decay experiments, while an unexpected 
small value of $\gA{0}$ was found from the EMC data:
The quarks contribute only a small fraction to the proton's spin, 
and $g_A(0)
$, therefore, attracted much attention \Ref{Anselmino}.

The investigations of the axial coupling constants by QCD sum rules 
have been done so far by the authors in 
Refs.[\ref{Belyaev83}-\ref{Belitsky}].
Belyaev and Kogan \Ref{Belyaev83} 
calculated $\gA{3}$ by considering a two-point correlation function of
nucleon currents in an external axial-vector field.
Ioffe \Ref{Ioffe98} also calculated $\gA{3}$ and $\gA{8}$ 
with the same correlation function.
Their method has some difficulties:
In the operator product expansion (OPE), there appear
new vacuum expectation values of 
quark-gluon composite operators induced by the external field.
In addition, the spectral function of the correlation function 
has double and single-pole terms.
The residue of the former is proportional to $g_{A}(0)$, 
and the latter corresponds to the transitions 
of a nucleon state to excited states 
through the interaction with the external field, but  
their residues are not known.
The authors used a method to obtain the values of $g_{A}(0)$ and the
residue of the single pole term from a $\chi^2$ fitting procedure,
and obtained a good agreement with experiment.

In this paper, we propose a new method to construct QCD sum rules for
the axial coupling constants from two-point correlation functions 
of the axial-vector currents in a one-nucleon state.
With the method, the correlation functions are expressed
by nucleon matrix elements of quark-gluon composite operators.
The spectral function has only a single pole term,
whose residue is related to $g_{A}(0)$.
As will be shown later, the axial coupling constants $g_{A}^{(3)}(0)$
and $g_{A}^{(8)}(0)$ are related 
to experimentally or theoretically
well-known quantities such as the $\pi$-$N$ and $K$-$N$ sigma-terms and
the moments of parton distributions.
For the singlet axial coupling constant  $\gA{0}$ 
we need to fully take into account the chiral anomaly,
but it seems to be difficult within the ordinary framework of 
QCD sum rules, and we leave it as a future work.


We first consider a correlation function of axial-vector currents:
\ben
\Pi^{(i)}_{\mu\nu}(q;P)
=i\int dx^4 {\rm e}^{iqx}\lgl T[j_{\mu 5}^{(i)}(x),j_{\nu 5}^{(i)}(0)]\rgl_N,
                                  \label{corfun}
\een
where the superscript $i$ is the $SU(3)_f$ index and $q^{\mu}\equiv(\omega,{\bf q})$.
In \eq{corfun}, the nucleon matrix element is defined by
\ben
\lgl\ldots \rgl_N \equiv {1\over2}\sum_S
\left[
\lgl N(PS)|\ldots |N(PS)\rgl-\lgl\ldots\rgl_0 \lgl N(PS)|N(PS)\rgl
\right],\nnbr
\een
where $P^{\mu}\equiv(E,{\bf P})$ is the nucleon momentum ($P^2=M^2$,
$M$ is the nucleon mass), $S$ the nucleon spin, $\lgl\ldots\rgl_0$ the vacuum 
expectation value, 
and the one-nucleon state to be normalized as $\lgl N(PS)|N(P'S')\rgl=(2\pi)^3 
\delta^{3}({\bf P}-{\bf P'})\delta_{SS'}$. 
The axial-vector currents are defined as
\bey
j_{\mu 5}^{(3)}(x)&=&\frac{1}{2}
\eta_{\mu\nu}\left[{\bar u}(x)\gamma^{\nu}\gamma_{5}u(x)-{\bar 
d}(x)\gamma^{\nu}\gamma_{5}d(x)\right],
                                              \label{jmu53} \\
j_{\mu 5}^{(8)}(x)&=&
\frac{1}{2\sqrt{3}}
\left[{\bar u}(x)\gamma_{\mu}\gamma_{5}u(x)+{\bar 
d}(x)\gamma_{\mu}\gamma_{5}d(x)
-2{\bar s}(x)\gamma_{\mu}\gamma_{5}s(x)\right],          \label{jmu58}
\eey
where $u$, $d$ and $s$ are the up, down and strange quark fields,
respectively.
In \eq{jmu53}, $\eta_{\mu\nu}\equiv q_{\mu}q_{\nu}/q^2-g_{\mu\nu}$
is introduced to make the current conserved and
suppress, simultaneously, the pion contribution to the current \Ref{yazaki}.

The Lehmann representation of the correlation function is given by
\ben
\Pi_{\mu\nu}(\omega,{\bf q};P)=\int^{\infty}_{-\infty}
d\omega'\frac{\rho_{\mu\nu}(\omega',{\bf 
q};P)}{\omega-\omega'},\label{Lehmann}
\een
where $\rho_{\mu\nu}$ is the spectral function defined by
\bey
\rho_{\mu\nu}(\omega,{\bf q};P)
&\equiv&-\frac{1}{\pi}{\rm Im}\Pi_{\mu\nu}(\omega+i\epsilon,{\bf q};P).
\eey
Following Refs.\Refs{Kondo93}{Kondo96}, 
we split $\Pi_{\mu\nu}$ into even and odd parts as 
$\Pi_{\mu\nu}(\omega)=\Pi_{\mu\nu}(\omega^2)_{\rm 
even}+\omega\Pi_{\mu\nu}(\omega^2)_{\rm odd}$,
and make a Borel transform on both sides of each Lehmann representation:
\bey
&&\widehat{B}[\Pi_{\mu\nu}(\omega^2,{\bf q};P)_{\rm even}]
=-\int^{\infty}_{-\infty}
d\omega'\omega' \exp\left(-\omega'^2/s\right)\rho_{\mu\nu}(\omega',{\bf q};P),
\label{BSR-even} \\
&&\widehat{B}[\Pi_{\mu\nu}(\omega^2,{\bf q};P)_{\rm odd}]
=-\int^{\infty}_{-\infty}
d\omega' \exp\left(-\omega'^2/s\right)\rho_{\mu\nu}(\omega',{\bf q};P), 
\label{BSR-odd}
\eey
where $s$ is a squared Borel mass 
and $\widehat{B}$ the Borel transformation:
\ben
\widehat{B}\equiv
\lim_{\stackrel{\stackrel{-\omega^2\rightarrow\infty}{n\rightarrow\infty}}
{\frac{-\omega^2}{n}=s}}
          \frac{(-\omega^2)^{n+1}}{n!}
          \left[ -\frac{d}{d(-\omega^{2})} \right]^n.
\een
In \eqs{BSR-even}{BSR-odd} the left hand 
sides are evaluated by OPE's, which give rise to the Borel transformed 
QCD sum rules.

Let us now consider the physical content of the spectral function.
Among the intermediate states of the spectral function, the lowest one is 
a one-nucleon state.
The continuum state consists of meson-nucleon states, excited nucleon states and 
so on.
There is an energy gap between the pole of the one-nucleon state and the 
threshold of the continuum states.

The contribution of the one-nucleon state to the spectral function is 
expressed as
\bey
& &\rho^{(i)}_{\mu\nu}(\omega,{\bf q};P)\cr
&=&-\frac{M}{E_{+}}\delta(\omega+E-E_{+})
\frac{1}{2}\sum_{S,S'}\langle N(PS)|\jmu5 |N(P_+S')\rangle\langle N(P_+S')|\jnu5 
|N(PS)\rangle\cr
& &+\frac{M}{E_{-}}\delta(\omega-E+E_{-})
\frac{1}{2}\sum_{S,S'}\langle N(PS)|\jnu5 |N(P_-S')\rangle\langle N(P_-S')|\jmu5 
|N(PS)\rangle,
\label{Pi-N}
\eey
where $P_{\pm}\equiv(E_{\pm},{\bf P}_{\pm})=(E\pm\omega,{\bf P}\pm{\bf q})$ 
are the four-momenta of the one-nucleon states. From the definitions of 
the axial coupling constants, the matrix elements in the right hand side of 
\eq{Pi-N} are written as
\bey
\lgl N(PS)|j_{\mu 5}^{(3)}(0)|N(P'S')\rgl&=&\eta_{\mu\nu}
{\bar u}(PS)(\lambda^{3}/2)\left[g_{A}^{(3)}(q^2)\gamma^{\nu}\gamma_5
\right]u(P'S'),                       \label{def-ga3}\\
\lgl N(PS)|j_{\mu 5}^{(8)}(0)|N(P'S')\rgl&=&
{\bar u}(PS)(\lambda^{8}/2)\left[g_{A}^{(8)}(q^2)\gamma_{\mu}\gamma_5
+h_{A}^{(8)}(q^2)
q_{\mu}\gamma_5 \right]u(P'S'),                       \label{def-ga8}
\eey
where $u(PS)$ is a Dirac spinor, $\lambda$ the usual Gell-Mann matrix and 
$q=P'-P$.
Note that the pseudo-coupling term disappears owing to $\eta_{\mu\nu}$
in \eq{def-ga3}.

The continuum contribution becomes small in the Borel transformed QCD 
sum rules,
since it is exponentially suppressed compared to the 
one-nucleon state because of the energy gap.
Therefore, it is allowed to use a rough model of the continuum:
The form of the continuum is approximated 
by the step function with the coefficient 
being the imaginary part of the asymptotic form 
of the correlation function in the 
OPE \Refs{SVZ1}{SVZ2}.
In the present case, however, the continuum contribution to the 
spectral function is absent within the approximation,
because from the definition of the correlation function the
perturbative part is subtracted.
This means that the continuum contribution may be very small
at least in the high energy region.
In the following we therefore neglect the continuum contribution 
to the spectral function.

Hereafter we consider the currents in which the Lorentz indices are
contracted, and the correlation function in the rest 
frame of the initial and final nucleon states, ${\bf P}=0$.
Thus we simplify our notation as follows: $\Pi(\omega,{\bf 
q})={\Pi_\mu}^\mu(\omega,{\bf q};M,{\bf 0})$, $\rho(\omega,{\bf 
q})={\rho_\mu}^\mu(\omega,{\bf q};M,{\bf 0})$. 
Then the spectral functions become 
\bey
\rho^{(3)}(\omega,{\bf q})
&=&-\left({1\over2}\right)^2\frac{1}{M\sqrt{M^2+|{\bf q}|^2}}
|\ga{3}|^2(q^2/2-2M^2)\nnbr
\\
& &\times
\left[\delta(\omega+M-\sqrt{M^2+|{\bf
q}|^2})-\delta(\omega-M+\sqrt{M^2+|{\bf q}|^2})\right], 
           \label{rho3-N} \\
\rho^{(8)}(\omega,{\bf q})
&=&-\left({1\over2\sqrt{3}}\right)^2\frac{1}{M\sqrt{M^2+|{\bf
q}|^2}}
\nnbr \\
& &\times\left[|\ga{8}|^2(q^2/2-3M^2)+\ga{8}\ha{8}\cdot
Mq^2-|\ha{8}|^2(q^4/4)\right]\nnbr\\ 
& &\times
\left[\delta(\omega+M-\sqrt{M^2+|{\bf
q}|^2})-\delta(\omega-M+\sqrt{M^2+|{\bf q}|^2})\right].
                    \label{rho8-N}
\eey
Because of crossing symmetry, 
\eqs{rho3-N}{rho8-N} are even functions of $\omega$, so that
the Lehmann representations of \eq{BSR-odd} is not necessary. 
From Eqs.(\ref{BSR-even}), (\ref{rho3-N}) and (\ref{rho8-N}) we obtain
\bey
\hat{B}\left[\Pi^{(3)}(\omega,{\bf q})_{\rm even}\right]
&=&\left({1\over2}\right)^2
\left(\frac{1}{M}-\frac{1}{\sqrt{M^2+|{\bf q}|^2}}\right)
{\rm exp}\left[-\left(\sqrt{M^2+|{\bf q}|^2}-M\right)^2/s\right]\nnbr\\
& &\times |g^{(3)}_{A}(q^2)|^2(q^2-4M^2),\label{triplet-N}\\
\hat{B}\left[\Pi^{(8)}(\omega,{\bf q})_{\rm even}\right]
&=&\left({1\over2\sqrt{3}}\right)^2
\left(\frac{1}{M}-\frac{1}{\sqrt{M^2+|{\bf q}|^2}}\right)
{\rm exp}\left[-\left(\sqrt{M^2+{\bf q}|^2}-M\right)^2/s\right]
\nnbr\\
&\times& \left[|g^{(8)}_{A}(q^2)|^2 (q^2-6 M^2)
+2M q^2 g^{(8)}_{A}(q^2) h^{(8)}_{A}(q^2)
-{q^4\over2} |h^{(8)}_{A}(q^2)|^2 \right].\label{octet-N}
\eey
We expand the right hand sides of \eqs{triplet-N}{octet-N} in powers of $\3mmtm$.
There is no constant term in the contribution of the one-nucleon state 
to the spectral function. 
The coefficients of $|{\bf q}|^2$ in \eqs{triplet-N}{octet-N}
are proportional to
$|g^{(3)}_{A}(0)|^2$ and $|g^{(8)}_{A}(0)|^2$, respectively.
Note that $h_{A}^{(8)}(0)$ contributes to higher order terms 
because $h_{A}^{(8)}(q^2)$ has no singularity at $q^2=0$.
Taking the first derivative with respect to $\3mmtm$
we obtain the desired QCD sum rules at $|{\bf q}|^2=0$ : 
\bey
\left.{\partial\over\partial|{\bf q}|^2} 
\hat{B}\left[\Pi^{(3)}(\omega^2,{\bf q})_{\rm even}\right]
\right|_{|{\bf q}|^2=0}
&=&-\left({1\over2}\right)^2\frac{2}{M}|g^{(3)}_{A}(0)|^2
, \label{BSR3}\\
\left.{\partial\over\partial|{\bf q}|^2} 
\hat{B}\left[\Pi^{(8)}(\omega^2,{\bf q})_{\rm even}\right]
\right|_{|{\bf q}|^2=0}
&=&-\left({1\over2\sqrt{3}}\right)^2\frac{3}{M}|g^{(8)}_{A}(0)|^2
. \label{BSR8}
\eey


Let us now turn to the OPE of $\Pi^{(3)}$ and $\Pi^{(8)}$.
In the OPE of \eq{corfun}, operators of the leading terms are of dimension 4. 
In this work we take into account the terms up to dimension 6.
The result for the iso-vector channel correlation function is 
in the following:
\bey
\Pi^{(3)}(q)&=&\left(\frac{1}{2}\right)^2\left[
\frac{6}{q^2}\left(\m{u}\N{u}+\m{d}\N{d}\right)
-\frac{1}{2q^2}\ggN 
\right. \nnbr \\
& &-\frac{8q^{\mu}q^{\nu}}{q^4}
\left(\gdN{u}+\gdN{d}\right) \nnbr \\
& &-\frac{6\pi\alpha_s}{q^4}
      \left\langle\left({\bar u}\gamma_{\mu}\lambda^{a}u
                       -{\bar d}\gamma_{\mu}\lambda^{a}d\right)^2
      \right\rangle_{N} \nnbr \\
& &+\frac{8\pi\alpha_s q^{\mu}q^{\nu}}{q^6}
      \left\langle{\cal S}\left({\bar u}\gamma_{\mu}\lambda^{a}u
                       -{\bar d}\gamma_{\mu}\lambda^{a}d
                       \right)
                          \left(\mu\rightarrow\nu\right)
      \right\rangle_{N} \nnbr \\
& &-\frac{4\pi\alpha_s}{3q^4}
      \left\langle\left({\bar u}\gamma_{\mu}\lambda^{a}u
                       +{\bar d}\gamma_{\mu}\lambda^{a}d\right)
                       \sum_{q=u,d,s}{\bar q}\gamma^{\mu}\lambda^{a}q
      \right\rangle_{N} \nnbr \\
& &+\frac{2\pi\alpha_s q^{\mu}q^{\nu}}{q^6}
      \left\langle{\cal S}\left({\bar u}\gamma_{\mu}\lambda^{a}u
                       +{\bar d}\gamma_{\mu}\lambda^{a}d
                       \right)
                       \sum_{q=u,d,s}{\bar q}\gamma_{\nu}\lambda^{a}q
      \right\rangle_{N} \nnbr \\
& &\left.+\frac{32 q^{\mu}q^{\nu}q^{\lambda}q^{\sigma}}{q^8}
      \left(\gdddN{u}+\gdddN{d}\right)\right].         \label{OPE3}
\eey
Similarly, the result for the 8th component of the SU(3) octet is given by
\bey
\Pi^{(8)}(q)&=&\left(\frac{1}{2\sqrt{3}}\right)^2\left[
\frac{10}{q^2}\left(\m{u}\N{u}+\m{d}\N{d}+4\m{s}\N{s}
\right)
-\frac{3}{2q^2}\ggN 
\right. \nnbr \\
& &-\frac{8q^{\mu}q^{\nu}}{q^4}
\left(\gdN{u}+\gdN{d}+4\gdN{s}\right) \nnbr \\
& &-\frac{6\pi\alpha_s}{q^4}
      \left\langle\left({\bar u}\gamma_{\mu}\lambda^{a}u
                       +{\bar d}\gamma_{\mu}\lambda^{a}d
                       -2{\bar s}\gamma_{\mu}\lambda^{a}s\right)^2
      \right\rangle_{N} \nnbr \\
& &+\frac{8\pi\alpha_s q^{\mu}q^{\nu}}{q^6}
      \left\langle{\cal S}\left({\bar u}\gamma_{\mu}\lambda^{a}u
                       +{\bar d}\gamma_{\mu}\lambda^{a}d
                       -2{\bar s}\gamma_{\mu}\lambda^{a}s\right)
                          \left(\mu\rightarrow\nu\right)
      \right\rangle_{N} \nnbr \\
& &-\frac{4\pi\alpha_s}{3q^4}
      \left\langle\left({\bar u}\gamma_{\mu}\lambda^{a}u
                       +{\bar d}\gamma_{\mu}\lambda^{a}d
                       +4{\bar s}\gamma_{\mu}\lambda^{a}s\right)
                       \sum_{q=u,d,s}{\bar q}\gamma^{\mu}\lambda^{a}q
      \right\rangle_{N} \nnbr \\
& &+\frac{2\pi\alpha_s q^{\mu}q^{\nu}}{q^6}
      \left\langle{\cal S}\left({\bar u}\gamma_{\mu}\lambda^{a}u
                       +{\bar d}\gamma_{\mu}\lambda^{a}d
                       +4{\bar s}\gamma_{\mu}\lambda^{a}s\right)
                       \sum_{q=u,d,s}{\bar q}\gamma_{\nu}\lambda^{a}q
      \right\rangle_{N} \nnbr \\
& &+\frac{32 q^{\mu}q^{\nu}q^{\lambda}q^{\sigma}}{q^8}
      \left(\gdddN{u}+\gdddN{d}\right. \nnbr \\
& &\qquad\qquad\qquad
             \left.\left.+4\gdddN{s}\right)\right],        \label{OPE8}
\eey
where $D_\mu$'s are covariant derivatives,
$G^2\equiv G^a_{\mu\nu}G^{a\mu\nu}$, and ${\cal S}$ denotes a symbol which
makes the operators symmetric and traceless with respect to the Lorentz
indices.

We now discuss about the nucleon matrix elements in \eqs{OPE3}{OPE8}.
It is  known well that $m_{q}\N{q}$ is related to the $\pi$-$N$ or
$K$-$N$ sigma-term as
\bey
&&(\m{u}+\m{d})(\N{u}+\N{d})=2\sgmtrm,\\
&&(\m{s}+\m{u})(\langle\bar ss\rangle_N+\langle\bar uu\rangle_N)
=2\Sigma_{KN}.
\eey
$\lgl(\alpha_s /\pi)G^2\rgl_{N}$ 
is expressed by the nucleon mass and $m_{q}\N{q}$ through the
QCD trace anomaly:
\ben
\ggN=-\frac{8}{9}\Big(M-\sum_{q=u,d,s} m_{q}\N{q}\Big).
\een
The matrix elements which contain covariant derivatives are related to 
the parton distributions as
\bey
&&\langle{\cal S}{\bar q}\gamma_{\mu_{1}}D_{\mu_{2}}\cdots
D_{\mu_{n}}q(\mu^2)\rangle_{N}
=(-i)^{n-1}A_{n}^{q}(\mu^2)T_{\mu_{1}\ldots \mu_{n}},
\eey
where $A_{n}(\mu^2)$ is the $n$-th moment of the parton distributions 
at scale $\mu^2$, 
and $T_{\mu_{1}\ldots \mu_{n}}={\cal S}\left[P_{\mu_{1}}\cdots 
P_{\mu_{n}}\right]$.
For the matrix elements of four quark operators,
we apply the factorization hypothesis:
In the vacuum, four quark condensates
are factorized by the hypothesis
which assumes that the vacuum contribution dominates in the
intermediate states: $\lgl{\cal O}_1 {\cal O}_2 \rgl_0 \approx \lgl{\cal O}_1
\rgl_0 \lgl{\cal O}_2 \rgl_0$ \Refs{SVZ1}{SVZ2}.
Similarly, for the nucleon matrix elements, we assume that the
contribution from one nucleon state dominates in the intermediate
states \Refs{Kondo93}{Kondo98}:
$\lgl{\cal O}_1 {\cal O}_2 \rgl_N\approx
\lgl{\cal O}_1 \rgl_N \lgl{\cal O}_2 \rgl_0 +\lgl{\cal O}_1 \rgl_0 \lgl{\cal 
O}_2 \rgl_N $. 
We apply this hypothesis to the following type of the
nucleon matrix elements,
which appear in \eqs{OPE3}{OPE8}: $\lgl{\bar q}_{f}\gamma_{\mu}\lambda^a q_{f}
{\bar q}_{f'}\gamma_{\nu}\lambda^a q_{f'}\rgl_{N}
=-(8/9)g_{\mu\nu}\lgl{\bar q}_{f}q_{f}\rgl_{0}
\lgl{\bar q}_{f}q_{f}\rgl_{N}\delta_{f,f'}$,
where $f$ and $f'$ are flavor indices.

We substitute \eqs{OPE3}{OPE8} into the left hand sides of \eqs{BSR3}{BSR8}, respectively. Averaging over the iso-spin states, we obtain the QCD sum 
rules for $|\gA{3}|^2$ and $|\gA{8}|^2$ as follows:
\bey
|\gA{3}|^2&=&-\frac{M}{2}\left\{
\frac{\sgmtrm}{s}\left[\frac{50}{9}+\frac{4}{9}{\m{s}\over\m{u}+\m{d}}\right]
-\frac{\Sigma_{KN}}{s}\left[\frac{8}{9}{\m{s}\over\m{s}+\m{u}}\right]
\right.\nnbr\\
&&-\frac{M}{s}\left[\frac{4}{9}+7(\mmnt{2}{u}+\mmnt{2}{d})\right] \nnbr \\
&&-\frac{4\pi\alpha_s\qqv}{s^2}
 \left[\frac{352}{27}\frac{\sgmtrm}{\m{u}+\m{d}}\right] 
+\left.\frac{15M^3}{s^2}\left[\mmnt{4}{u}+\mmnt{4}{d}\right]\right\},
\label{QSR3}\\
|\gA{8}|^2&=&-\frac{M}{3}\left\{
\frac{\sgmtrm}{s}\left[\frac{26}{3}-\frac{116}{3}{\m{s}\over\m{u}+\m{d}}\right]
+\frac{\Sigma_{KN}}{s}\left[\frac{232}{3}{\m{s}\over\m{s}+\m{u}}\right]
\right. \nnbr \\
&&
+\frac{M}{s}\left[\frac{4}{3}-7\left(\mmnt{2}{u}
+\mmnt{2}{d}+4\mmnt{2}{s}\right)\right]
\nnbr \\
&&
-\frac{4\pi\alpha_s\qqv}{s^2}
 \left[\frac{352}{27}\frac{\sgmtrm}{\m{u}+\m{d}}\right]
-\frac{4\pi\alpha_s\ssv}{s^2}
\left[\frac{704}{27}\left(\frac{2\Sigma_{KN}}{\m{s}+\m{u}}-\frac{\sgmtrm}{\m{u}+
\m{d}}\right)\right]\cr
&&\left.
+\frac{15M^3}{s^2}\left[\mmnt{4}{u}+\mmnt{4}{d}+4\mmnt{4}{s}\right]
\right\},  \label{QSR8} 
\eey
respectively, where $\qqv\equiv \uuv=\ddv$. 
In \eqs{QSR3}{QSR8} we assume $m_u=m_d$. From these equations, we 
find that the axial coupling
constants are 
related to the sigma-terms and the moments of parton distributions.
Since the sigma-terms and the moments
are well known, we can estimate $|\gA{3}|$ and $|\gA{8}|$.

\begin{figure}[t]
\includegraphics{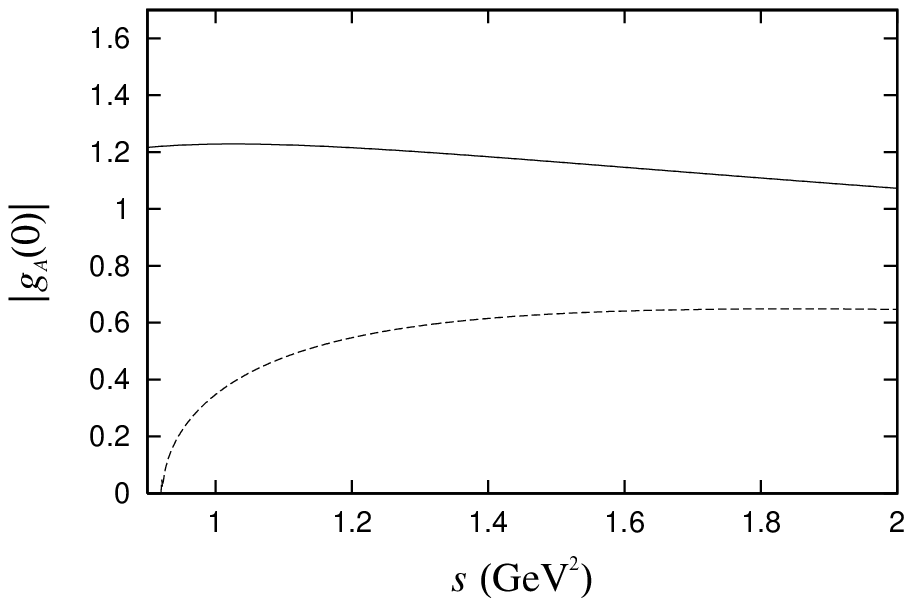}
\vspace{.5cm}
\\
\begin{center}
{{\bf Fig.1}. The squared Borel mass $s$ dependence of $|\gA{3}|$ in
\eq{QSR3} and $|\gA{8}|$ in \eq{QSR8}. The solid line corresponds to $|\gA{3}|$ and 
the dashed line to $|\gA{8}|$.}
\end{center}
\end{figure}

We show in Fig.1 the squared Borel mass $s$ dependence of $|\gA{3}|$ in
\eq{QSR3} and $|\gA{8}|$ in \eq{QSR8}.
In plotting the curve in Fig.1, we used the following values
of the constants in the OPE.
The $\pi$-$N$ sigma-term is taken from Ref.[\ref{GLM}], which are $\sgmtrm 
=45\,{\rm MeV}$. 
The quark masses are taken to be $\m{u}=\m{d}=7\,{\rm MeV}$, 
$\m{s}=110\,{\rm MeV}$ \Ref{yazaki}.

Using the above values and the ratio 
$2\N{s}/(\N{u}+\N{d})=0.2$ given in Ref.[\ref{GLM}], we can calculate the $K$-$N$ sigma term averaged over the iso-spin states 
and the result is $\Sigma_{KN}=226\,{\rm MeV}$. 
Following Ref.\Ref{HL} and adopting the LO scheme in Ref.\Ref{Gluck}, we calculated the moments of parton
distributions: 
$\mmntQSR{2}{u}+\mmntQSR{2}{d}= 1.1$, 
$\mmntQSR{4}{u}+\mmntQSR{4}{d}= 0.13$,
$\mmntQSR{2}{s}= 0.03$, $\mmntQSR{4}{s}= 0.002$.
The vacuum condensates are taken from Ref.\Ref{yazaki}, 
which are $\qqv=(-225\,{\rm 
MeV})^3$ and $\ssv=0.8\qqv$.
Among the above constants in the OPE, the most dominant contribution
comes from $\mmntQSR{2}{u}+\mmntQSR{2}{d}$.

From Fig.1 we see good stability of the $s$ dependence for both $\gA{3}$ 
and $\gA{8}$.
We see that the variations of the curves in a larger $s$ region are small 
in spite of not including the continuum contribution.
This fact implies that the continuum contribution is suppressed,
and the spectral function 
is allowed to be approximated by the lowest state.

Our estimations of $|\gA{3}|$ and $|\gA{8}|$ taken from the stabilized region
are
\ben
|\gA{3}|\simeq 1.2,\quad |\gA{8}|\simeq 0.6.
\een
The obtained value of $|\gA{3}|$ gives a good agreement 
with the world average
$\gA{3}=1.260\pm 0.002$ \Ref{exp-ga3}.
The value of $|\gA{8}|$ is also close to $\gA{8}=0.59\pm 0.02$
\Ref{exp-ga8} found from the data on the baryon octet $\beta$-decays 
under the assumption of the SU(3) flavor symmetry.


In summary, we have considered two-point correlation functions
of axial-vector currents in one-nucleon state and found that the 
lowest state in the 
spectral function of the correlation function is expressed by the axial coupling 
constant. 
We have calculated the iso-vector and the 8th component of the SU(3) octet
axial coupling constants in the framework of QCD sum rules.
The results show that the axial coupling constants are expressed by the
nucleon matrix elements of quark and gluon operators 
which are related to the sigma-terms and the moments of parton distributions.
Since the nucleon matrix elements are known well experimentally 
or theoretically,
the axial coupling constants are calculated   with a small ambiguity, and
the obtained results are in a good agreement with experiment.

Finally, we mention about the singlet axial coupling constant,
$\gA{0}$, which is considered to be the nucleon spin carried by quarks.
The unexpected small value of $\gA{0}$ found by EMC
has raised a number of understanding 
of the dynamics of the nucleon spin \Ref{Anselmino}.
Within the same framework as those for $\gA{3}$ and $\gA{8}$ ,
we find that the calculated $\gA{0}$ is about $0.8$, which is
 not so small as the value found by EMC.
In this calculation, however, the effect of the chiral anomaly 
is not taken into account.

The authors in Ref.\Ref{Belitsky}
have calculated $\gA{0}$ fully taking into account the anomaly relation.
They have considered a three-point function of nucleon 
interpolating fields and the divergence of the singlet axial-vector current.
The form factors, $g_{A}^{(0)}(q^2)$, are related to 
the vacuum condensates of the 
quark-gluon composite operators through the double dispersion relation. 
To know $g_{A}^{(0)}(q^2)$ at $q^2=0$ one must evaluate the correlation
function at the zero momentum.
Although the method to evaluate it is known \Ref{Balitsky}, 
it involves large uncertainty.
The calculation of $\gA{0}$ in our approach by taking into account 
the chiral anomaly will be reported elsewhere \Ref{ga0}.   

The authors (T. N. and Y. K.) would like to thank 
M. Oka, A. Hosaka and M. Takizawa for fruitful discussions.
They also wish to acknowledge helpful discussions with O. Morimatsu
and H. Terazawa.
Y. K. thanks the theory group of IPNS(Tanashi) for the hospitality.

\newpage
\baselineskip 24pt
\begin{center}
{\bf References}
\end{center}
\def\labelenumi{[\theenumi]}
\def\Ref#1{[\ref{#1}]}
\def\Refs#1#2{[\ref{#1},\ref{#2}]}
\def\npb#1#2#3{{Nucl. Phys.\,}{\bf B{#1}}\,(#3), #2}
\def\npa#1#2#3{{Nucl. Phys.\,}{\bf A{#1}}\,(#3),#2}
\def\np#1#2#3{{Nucl. Phys.\,}{\bf{#1}}\,(#3),#2}
\def\plb#1#2#3{{Phys. Lett.\,}{\bf B{#1}}\,(#3),#2}
\def\prl#1#2#3{{Phys. Rev. Lett.\,}{\bf{#1}}\,(#3),#2}
\def\prd#1#2#3{{Phys. Rev.\,}{\bf D{#1}}\,(#3),#2}
\def\prc#1#2#3{{Phys. Rev.\,}{\bf C{#1}}\,(#3),#2}
\def\pr#1#2#3{{Phys. Rev.\,}{\bf{#1}}\,(#3),#2}
\def\ap#1#2#3{{Ann. Phys.\,}{\bf{#1}}\,(#3),#2}
\def\prep#1#2#3{{Phys. Reports\,}{\bf{#1}}\,(#3),#2}
\def\rmp#1#2#3{{Rev. Mod. Phys.\,}{\bf{#1}}\,(#3),#2}
\def\cmp#1#2#3{{Comm. Math. Phys.\,}{\bf{#1}}\,(#3),#2}
\def\ptp#1#2#3{{Prog. Theor. Phys.\,}{\bf{#1}}\,(#3),#2}
\def\ib#1#2#3{{\it ibid.\,}{\bf{#1}}\,(#3),#2}
\def\zsc#1#2#3{{Z. Phys. \,}{\bf C{#1}}\,(#3),#2}
\def\zsa#1#2#3{{Z. Phys. \,}{\bf A{#1}}\,(#3),#2}
\def\intj#1#2#3{{Int. J. Mod. Phys.\,}{\bf A{#1}}\,(#3),#2}
\def\sjnp#1#2#3{{Sov. J. Nucl. Phys.\,}{\bf #1}\,(#3),#2}
\def\pan#1#2#3{{Phys. Atom. Nucl.\,}{\bf #1}\,(#3),#2}
\def\app#1#2#3{{Acta. Phys. Pol.\,}{\bf #1}\,(#3),#2}

\def\etal{{\it et al.}}
\begin{enumerate}

\divide\baselineskip by 4
\multiply\baselineskip by 3

\item \label{Anselmino}
M. Anselmino, A. Efremov and E. Leader, \prep{261}{1}{1995}
\item\label{Belyaev83}
V. M. Belyaev and Ya. I. Kogan, JETP Lett.\,{\bf 37}\,(1983),20.
\item \label{Belyaev85}
V. M. Belyaev, B. L. Ioffe and Ya. I. Kogan, \plb{151}{290}{1985}.
\item \label{Ioffe92}
B. L. Ioffe and A. Yu. Khodzhamiryan, \sjnp{55}{1701}{1992}
\item \label{Ioffe98}
B. L. Ioffe, talk given at St.Petersburg Winter School on Theoretical
Physics, hep-ph/9804238.
\item \label{Belitsky}
A. V. Belitsky and O. V. Teryaev, \pan{60}{455}{1997}
\item \label{yazaki}
L. J. Reinders, H. R. Rubinstein, and S. Yazaki,
\prep{127}{1}{1985}.
\item \label{Kondo93}
Y. Kondo and O. Morimatsu, \prl{2855}{71}{1993}.
\item \label{Kondo96}
Y. Kondo, O. Morimatsu and Y. Nishino, \prc{53}{1927}{1991}.
\item \label{SVZ1} M. A. Shifman, A. I. Vainshtein and~V. I. Zakharov, 
\npb{147}{385}{1979}.
\item \label{SVZ2} M. A. Shifman, A. I. Vainshtein and~V. I. Zakharov,
\npb{147}{448}{1979}.
\item \label{Kondo98}
Y. Kondo and O. Morimatsu, \ptp{1}{100}{1998}.
\item \label{GLM} J. Gasser, H. Leutwyler and M. E. Sainio,
\plb{253}{252}{1991}.
\item \label{HL}
T. Hatsuda and S. H. Lee, \prc{34}{R46}{1992}.
\item \label{Gluck} M. Gl\"uck, E. Reya and A. Vogt,
\zsc{53}{127}{1992}.
\item \label{exp-ga3}
R. M. Barnett et al., Particle Data Group, \prd{54}{1}{1996}.
\item \label{exp-ga8}
S. Y. Hsueh et al., \prd{38}{2056}{1988}.
\item \label{Balitsky}
Ya. Ya. Balitsky, A. V. Kolesnichenko and A. V. Yung,
\sjnp{41}{178}{1985}.
\item \label{ga0}
T. Nishikawa, S. Saito and Y. Kondo, in preparation.

\end{enumerate}

\end{document}